\documentclass[%
 aip,
 amsmath,amssymb,
 reprint,%
]{revtex4-1}

\usepackage{graphicx}% Include figure files
\usepackage{dcolumn}% Align table columns on decimal point
\usepackage{bm}% bold math
\usepackage[utf8]{inputenc}
\usepackage[T1]{fontenc}
\usepackage{mathptmx}
\usepackage{etoolbox}

\makeatletter
\def\@email#1#2{%
 \endgroup
 \patchcmd{\titleblock@produce}
  {\frontmatter@RRAPformat}
  {\frontmatter@RRAPformat{\produce@RRAP{*#1\href{mailto:#2}{#2}}}\frontmatter@RRAPformat}
  {}{}
}%
\makeatother
\begin{document}

\preprint{AIP/123-QED}

\title[A Novel Approach to Interface High-Q Fabry-Pérot Resonators with Photonic Circuits]{A Novel Approach to Interface High-Q Fabry-Pérot Resonators with Photonic Circuits}

\author{Haotian Cheng}
 \email{haotian.cheng@yale.edu}
 \affiliation{Department of Applied Physics, Yale University, New Haven, CT 06520, USA}
\author{Naijun Jin}
 \affiliation{Department of Applied Physics, Yale University, New Haven, CT 06520, USA}
\author{Zhaowei Dai}
 \affiliation{Department of Applied Physics, Yale University, New Haven, CT 06520, USA}
\author{Chao Xiang}
\affiliation{Department of Electrical and Computer Engineering, University of California, Santa Barbara, Santa Barbara, CA 93106, USA}
\author{Joel Guo}
\affiliation{Department of Electrical and Computer Engineering, University of California, Santa Barbara, Santa Barbara, CA 93106, USA}
\author{Yishu Zhou}
 \affiliation{Department of Applied Physics, Yale University, New Haven, CT 06520, USA}
\author{Scott A. Diddams}
 \affiliation{National Institute of Standards and Technology, 325 Broadway, Boulder, CO 80305, USA}
 \affiliation{Department of Physics, University of Colorado Boulder, 440 UCB Boulder, CO 80309, USA}
 \affiliation{Electrical Computer \& Energy Engineering, University of Colorado, Boulder, CO 80309, USA}
\author{Franklyn Quinlan}
 \affiliation{National Institute of Standards and Technology, 325 Broadway, Boulder, CO 80305, USA}
\author{John Bowers}
 \affiliation{Department of Electrical and Computer Engineering, University of California, Santa Barbara, Santa Barbara, CA 93106, USA}
\author{Owen Miller}
 \affiliation{Department of Applied Physics, Yale University, New Haven, CT 06520, USA}
\author{Peter Rakich}
 \email{peter.rakich@yale.edu}
 \affiliation{Department of Applied Physics, Yale University, New Haven, CT 06520, USA}
\date{\today}

\begin{abstract}
The unique benefits of Fabry-Pérot resonators as frequency-stable reference cavities and as an efficient interface between atoms and photons make them an indispensable resource for emerging photonic technologies. To bring these performance benefits to next-generation communications, computation, and timekeeping systems, it will be necessary to develop strategies to integrate compact Fabry-Pérot resonators with photonic integrated circuits. In this paper, we demonstrate a novel reflection cancellation circuit that utilizes a numerically optimized multi-port polarization-splitting grating coupler to efficiently interface high-finesse Fabry-Pérot resonators with a silicon photonic circuit. This circuit interface produces spatial separation of the incident and reflected waves, as required for on-chip Pound-Drever-Hall frequency locking, while also suppressing unwanted back reflections from the Fabry-Pérot resonator. Using inverse design principles, we design and fabricate a polarization-splitting grating coupler that achieves 55\% coupling efficiency. This design realizes an insertion loss of 5.8 dB for the circuit interface and more than 9 dB of back reflection suppression, and we demonstrate the versatility of this system by using it to interface several reflective off-chip devices.
\end{abstract}

\maketitle

\section{\label{sec:level1}Introduction}

High-finesse Fabry-Pérot cavities are unmatched in their ability to deliver high frequency stability, quality factors, and power handling, making them indispensable for a range of applications \cite{thompson_observation_1992,reiserer_cavity-based_2015,mckeever_deterministic_2004,colombe_strong_2007, matei_15um_2017, fortier_generation_2011}. To build next-generation quantum communications, computation, and timekeeping systems\cite{kudelin2023photonic}, it will be necessary to bring these performance advantages to compact, integrated platforms \cite{newman_architecture_2019,maurice_miniaturized_2020,spencer_optical-frequency_2018}. Using new wafer-scale fabrication techniques, it is now possible to make arrays of high-finesse (>$10^6$) Fabry-Pérot resonators \cite{jin2022micro,jin2022microfabrication} that have been used to create sub-Hz linewidth lasers \cite{guo2022chip} and low-noise oscillators \cite{mclemore2022miniaturizing}. However, to harness these performance advantages in next-generation integrated photonic circuits, we also require strategies to efficiently interface Fabry–Pérot cavities with photonic circuits\cite{kudelin2023photonic}.

The optimal strategy for integration of Fabry-Pérot resonators depends heavily on the intended use case. When a Fabry–Pérot resonator is used as a stable frequency reference for high-performance laser systems and optical clocks, the frequency of the cavity is typically measured using the Pound-Drever-Hall (PDH) locking technique; in this case, laser light reflected from the resonator must be separated from the incident wave and detected with high efficiency to obtain a low noise error signal for feedback stabilization of the laser frequency. Hence, implementation of on-chip PDH locking requires an integration strategy that maps incident and reflected waves to distinct ports of an optical system, permitting direct detection of the reflected wave\cite{kudelin2023photonic}. Ideally, this same photonic interface would also protect the laser from the frequency-destabilizing effects of back-scattered light by suppressing back reflections from the Fabry-Pérot resonator\cite{tkach1986regimes}.   
\begin{figure*}
\centering\includegraphics[width=13cm]{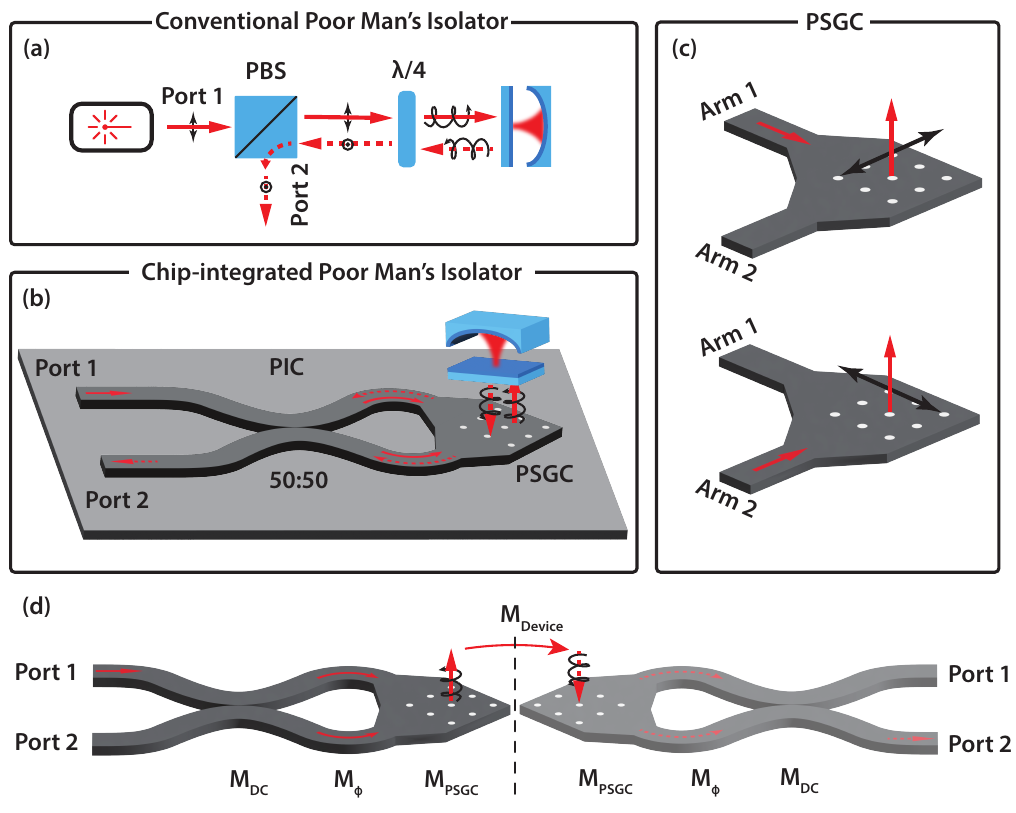}
\caption{(a) Depiction of the traditional setup of a poor man's isolator mechanism in free space, where PBS denotes a Polarizing Beam Splitter. (b) Illustration of the circuit interface, which is modeled after the poor man's isolator, that incorporates a the Polarization Splitting Grating Coupler (PSGC). PIC: photonic integrated circuits. (c) Schematic of PSGC. Light input from Arm 1 will be scattered into p-polarized light in free space and that from Arm 2 will be scattered into s-polarized light in free space. (d) Transfer matrix schematic of the system.}
\end{figure*}
An optical circulator is a natural solution to this problem, as it maps incident and reflected waves to distinct optical ports while offering some protection from back reflection. However, the fabrication of isolators and circulators on photonic chips poses a significant challenge due to the incompatibility of requisite magneto-optic materials\cite{bi2011chip,stadler2013integrated,du2018monolithic,pintus2019broadband} with CMOS foundries. To address this challenge, a variety of non-magnetic isolators\cite{kittlaus2021electrically,tian2021magnetic,zhou2022intermodal} and circulators\cite{herrmann2022mirror} have been demonstrated, which use time modulation to produce non-reciprocal response. However, since these non-magnetic isolators and circulators are complex and can consume a substantial amount of power, they are not suitable for all applications. 

New strategies for passive reflection cancellation could eliminate the need for isolators and circulators in many instances, offering a path to simpler and more power-efficient integrated photonic circuits. One such system that is widely used in free-space optics, colloquially referred to as the poor man’s isolator, uses a quarter-wave plate and a polarizing beam splitter to separate the incident and reflected optical waves. This system, pictured in Fig. 1(a), is frequently used instead of an optical circulator to implement PDH locking, since it offers lower losses and smaller back reflections. Hence, photonic circuit implementation of such systems could serve as a practical and efficient interface between Fabry-Pérot cavities and other free-space systems.

In this paper, we demonstrate a novel reflection cancellation circuit to efficiently interface high-finesse Fabry-Pérot resonators with a silicon photonic circuit. This system, whose operating principle is modeled after the poor man’s isolator, is comprised of an interferometer that interfaces to two separate ports of an optimized polarization splitting grating coupler (PSGC) device \cite{zaoui2013cmos,streshinsky2013compact,verslegers2014design,hu2022integrated}. Light entering Port 1 of the interferometer is reflected from a fiber-coupled Fabry-Pérot resonator before exiting Port 2 of the interferometer, yielding spatial separation of the incident and reflected waves as required for on-chip PDH locking. Using inverse design principles \cite{miller2012photonic,molesky2018inverse} to optimize the 2D grating structure, we demonstrate a peak fiber-to-chip coupling efficiency of 55\%, yielding 5.8~dB of loss in a double-pass configuration of the on-chip interface. Interferometric cancellation of reflections produced by this system also yields >9~dB of back-reflection suppression, which could help to protect an on-chip laser source from unwanted back reflections. 
 
Since the degree of back-reflection cancellation was limited only by the imprecision of the splitting ratio of a directional coupler, much higher (>30 dB) back-reflection suppression ratios should be possible with further refinements.   

\section{System Concept}
\begin{figure*}
\centering\includegraphics[width=13cm]{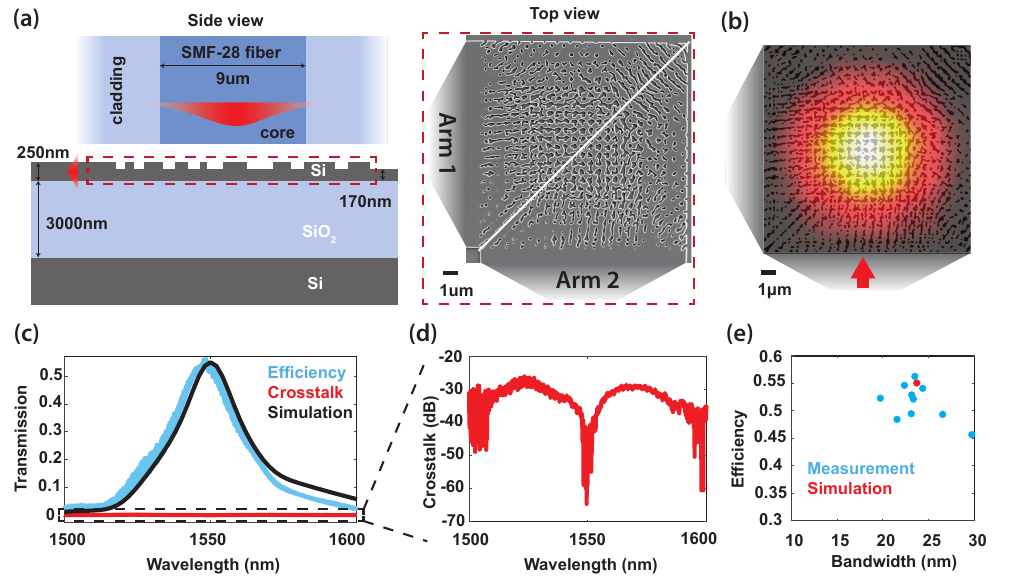}
\caption{(a) Side view: The layered structure of the PSGC is presented, with a single mode optical fiber positioned perpendicularly to the PSGC. Top view: SEM image of the PSGC. Pattern is symmetrical over the white line. (b) FDTD simulation showcases the scattered field 3 {\textmu}m above the Si device layer, prompted by an incident waveguide mode from bottom port. (c) Empirical measurement of the coupling efficiency and crosstalk of the PSGC. Efficiency is characterized as the transmission from Arm 1 to the fiber, while crosstalk is defined as the transfer from Arm 1 to Arm 2. (d) Zoom in of the crosstalk data presented in (c). (e) Scatter plot indicating the performance among ten distinct PSGCs fabricated on the same chip.}
\end{figure*}
The reflection cancellation circuit that we use to interface a high-finesse Fabry-Pérot resonator with a silicon photonic circuit is illustrated in Fig. 1. 
The operating principle of photonic circuit interface (Fig. 1(b)) closely mirrors the free-space implementation of the poor man’s isolator shown in Fig. 1(a). In the free-space implementation (Fig. 1(a)) p-polarized light entering Port 1 passes through the Polarizing Beam Splitter (PBS) and is subsequently conversion into right-handed circularly polarized light after it traverses the Quarter Wave Plate (QWP). Upon reflection from the Fabry-Pérot resonator, this right-handed circularly polarized wave is converted to left-handed circularly polarized wave. After traversing the QWP a second time, the left-handed circular polarized wave is converted to an s-polarized wave and is subsequently reflected by the PBS to exit Port 2 of the system. 
Hence, this system protects the laser source from back reflections while yielding spatial separation of the reflected wave as required for on-chip PDH locking. While we use an Fabry-Pérot resonator in this system demonstration, this scheme is applicable to any component with a polarization-independent reflection response.

The circuit implementation of this poor man's isolator system, seen in Fig. 1(b), is comprised of a balanced interferometer that incorporates an optimized polarization-splitting grating coupler (PSGC) device. 
In close analogy to the free-space polarizing beamsplitter, the PSGC maps orthogonally polarized waves into separate output waveguide arms. The 1(2) input arm of the PSGC couples to p-polarized (s-polarized) free-space beams that are emitted perpendicular to the grating coupler (Fig. 1(c)).
Light entering Port 1 of this interferometer is split between two waveguides by a 50/50 directional coupler before coupling to the PSGC. Since the directional coupler induces a $\pi/2$ phase difference between the two waveguides (each having identical path lengths), right-handed circularly polarized light is emitted from the PSGC.  Upon reflection from the Fabry-Pérot resonator (which is assumed to have a polarization-independent reflection response), the incident right-handed circular polarization is converted into a left-handed circular polarized wave before entering the PSGC for a second time. This left-handed circular polarized wave is projected into orthogonal linear polarizations by the PSGC, meaning that the waves exiting the PSGC now have a $-\pi/2$ phase difference. This phase difference causes these two reflected waves to combine within the 50/50 directional coupler, such that all of the reflected light exits Port 2 of the interferometer. Hence, this circuit interface yields spatial separation of the incident and reflected waves (as required for on-chip PDH locking) while also producing interferometric cancellation of unwanted back-reflections from Port~1.

Next, we use the transfer matrix formalism to analyze the response of this circuit interface. By mapping the bi-directional two-port system of Fig. 1(b) onto an equivalent unidirectional two-port system of Fig. 1(d), we can readily use $2\times2$ transfer matrices to analyze the system response. Denoting the incident waves as $\bigl[\begin{smallmatrix} a_1\\a_2\end{smallmatrix}\bigr]$ and the out-going waves as  $\bigl[\begin{smallmatrix} b_1\\b_2\end{smallmatrix}\bigr]$, we can find the system response by propagating the input waves through transfer matrices associated with each component of the system, as follows %Assuming that light is only 
\begin{figure*}
\centering\includegraphics[width=13cm]{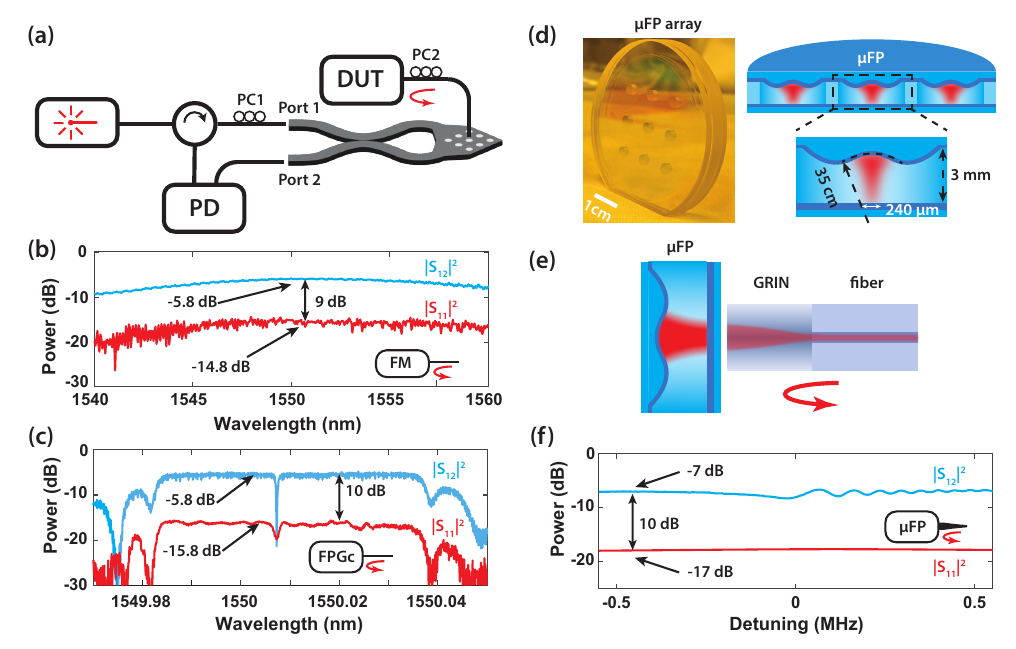}
\caption{On-chip poor man's isolator demonstration. (a) Measurement setup for the characterization of the reflection cancellation circuit. DUT: device under test, PD: photo detector. (b-c) Measurement results with fiber mirror (FM) and fiber Bragg grating cavity (FBGC) as DUT. (d) {\textmu}FP array. (e) Fiber GRIN lens collimator assembled to match {\textmu}FP cavity mode. (f) Measurement results with fiber-GRIN-{\textmu}FP as DUT.}
\end{figure*}

\begin{equation}
\bigl[\begin{smallmatrix}
b_1\\
b_2
\end{smallmatrix}\bigr] = \boldsymbol{M_{DC}}\boldsymbol{M_{\phi}}\boldsymbol{M_{PSGC}}\boldsymbol{M_{Device}}\boldsymbol{M_{PSGC}}\boldsymbol{M_{\phi}}\boldsymbol{M_{DC}} \bigl[\begin{smallmatrix}
a_1\\
a_2
\end{smallmatrix}\bigr].
\end{equation}

\noindent Here, $\boldsymbol{M_{DC}}\!=\!\bigl[\begin{smallmatrix}
r & -i\mu\\
-i\mu & r
\end{smallmatrix}\bigr]$, $\boldsymbol{M_{\phi}}\!=\!\bigl[\begin{smallmatrix}
1 & 0\\
0 & e^{i\phi}
\end{smallmatrix}\bigr]$, $\boldsymbol{M_{PSGC}}\!=\!\bigl[\begin{smallmatrix}
\alpha & 0\\
0 & \alpha
\end{smallmatrix}\bigr]$, and $ \boldsymbol{M_{Device}}\!\!=\!\!\bigl[\begin{smallmatrix}
\Tilde{R_1} & 0\\
0 & \Tilde{R_2}
\end{smallmatrix}\bigr]$ are transfer matrices that describe the response of the directional coupler, differential phase delay, PSGC and reflective device, respectively.

Here, $\mu$ and $r = \sqrt{1-\mu^2}$ are splitting coefficient of the directional coupler, $\phi$ is the phase imbalance between two waveguide segments, $\alpha$ is the coupling coefficient for PSGC, and $\Tilde{R_1}, \Tilde{R_2}$ are complex reflection coefficients for the external device (e.g., Fabry-Pérot resonator) for s- and p-polarized light. 

Using this model, we make some basic observations about this reflection cancellation scheme. Assuming that light is only injected into Port~1 (i.e., $a_2 = 0$), we find reflected wave amplitudes $b_1 = a_1\alpha^2 (\Tilde{R_1}r^2-\Tilde{R_2}\mu^2e^{2i\phi})$ and $b_2 = -a_1\alpha^2r\mu(\Tilde{R_1}+\Tilde{R_2}e^{2i\phi})$,  

from the above transfer matrix model. From these expressions, we see that it is possible to null $b_1$ even in cases when $\Tilde{R_1}\neq\Tilde{R_2}$. However, through experimental studies, we are interested in the case when the reflection response is identical for both polarizations. Assuming that $\Tilde{R_1}=\Tilde{R_2}$ in the case of 50\% power splitting ratio ($r=\mu=1/\sqrt2$), we see that it is possible to null $b_1$ while directing all of the output light $b_2$ in the case when $\phi=0$. Note, however, that the backreflection suppression ratio, $|\frac{b_2}{b_1}|^2=\frac{4\mu^2(1-\mu^2)}{|1-2\mu^2|^2}$, is sensitive to the splitting coefficient $\mu$. For example, if $\mu$ deviates from $1/\sqrt2$ by 14\% due to fabrication errors, the backreflection suppression ratio will decrease to 10 dB from perfect cancellation. Hence, tight control of the power splitting ratio is necessary to obtain a high degree of back-reflection cancellation.

\section{Results}

To realize this circuit interface, we fabricate a silicon photonic circuit from a Silicon-On-Insulator (SOI) wafer having a 250 nm thick silicon layer and a 3 micron silica under cladding. We use e-beam lithography and a reactive-ion etch (etch depth of 80 nm) to define both waveguide and grating structures, as seen in Fig. 2(a). To accurately define the desired structure during e-beam exposure, we implemented proximity effect correction \cite{owen1983proximity}, and through dose tests using hydrogen silsesquioxane (HSQ), an e-beam dose of  $1050~\rm{\mu C/cm}^2$ yielded the optimal performance. 
For details of the fabrication process, see \cite{otterstrom2018silicon}. The PSGC device is a key component of this system, as the efficiency of circuit interfaces hinges on the performance of this grating coupler.

To optimize the efficiency of the PSGC, we employ an inverse design algorithm\cite{miller2012photonic} using the LUMOPT package with Lumerical FDTD software\cite{NIST_DIS}. Through the design of the grating coupler, the parameter space of numerical optimization is reduced by imposing a mirror symmetry plane, indicated by the diagonal white line in Fig. 2(b). We modified the LUMOPT package in order to impose such mirror symmetry. This symmetry plane ensures that the grating coupler produces the same scattering response when excited from either input port. The PSGC structure (seen in Fig. 2(a-b)) is designed for vertical coupling to SMF-28 optical fiber, with a Gaussian beam waist radius of 5.2 {\textmu}m. Fig. 2(b) shows the Finite-Difference Time-Domain (FDTD) simulation of the Gaussian field profile produced by the PSGC device layer when illuminated by a TE-like guided optical mode.

To evaluate the optical performance of the fabricated PSGC structures, a cleaved optical fiber (SMF-28) is vertically aligned with the PSGC to enable fiber-to-chip coupling efficiency and cross-talk measurements. 
The optical measurements seen in Fig. 2(c), reveal fiber-to-chip coupling efficiency of 55\% (2.6 dB) at a wavelength of 1550 nm, over a 23 nm bandwidth. The crosstalk, defined as the direct coupling from Arm 1 to Arm 2, displays a contrast exceeding 60 dB at 1550 nm. Fig. 2(e) shows the measured peak efficiency and 3 dB bandwidth of 10 gratings fabricated on the same chip, revealing high efficiency (>50\%) and broad operational bandwidth (>20nm) for the majority of fabricated gratings. Adding reflective elements underneath PSGC will further help to increase to efficiency to nearly 100\%, avoiding light scattered into substrate\cite{luo2018low}.

Next, we use this optimized PSGC device to demonstrate the proposed reflection cancellation circuit, and we use this system to interface an off-chip Fabry-Pérot resonator with our circuit. Fig. 3(a) shows a schematic of the experimental apparatus that is used to assess the performance of the reflection cancellation circuit. The fiber circulator facilitates the measurement of system reflection, enabling the characterization of the back-reflection suppression ratio. Through these studies, we use a segment of SMF-28 fiber as a fiber umbilical to flexibly interface the grating coupler to different reflective devices. This fiber umbilical contains a polarization controller (PC2) that compensates for any polarization distortion occurring within the fiber. The state of the polarization controller is chosen to ensure that the fiber umbilical produces a Jones matrix of identity (i.e., doesn't alter the polarization state). Hence, the fiber umbilical is a convenient way to couple to different devices under test (DUT) to evaluate the performance of this circuit interface in various scenarios.

We begin by using a commercial fiber mirror (FM) as a reference device (Thorlabs P5-SMF28ER-P01-1) to examine the performance of the reflection cancellation circuit.  The reflection and transmission response produced by the fiber mirror are shown in Fig. 3(b). 
Power transmission from Port 1 to Port 2, denoted as $|S_{12}|^2$, reveals a signal attenuation of -5.8~dB. Since the light exiting Port 2 passes through the grating coupler twice, with 2.6~dB of loss per pass, and the fiber mirror has a typical reflection loss of $~0.6$~dB, the total transmission loss of 5.8~dB is in good agreement with the anticipated transmission loss of $~5.8$~dB. Measurements of the reflection response from Port~1, denoted as $|S_{11}|^2$, reveal a back-reflection of -14.8 dB from a fiber mirror with near-unity back-reflection efficiency. Since the reflection efficiency, $|S_{11}|^2$, is 9~dB lower than the transmission efficiency, these measurements reveal a back-reflection suppression ratio of 9~dB, validating the operating principle of this reflection cancellation circuit.

\begin{figure}
\centering\includegraphics[width=5.5cm]{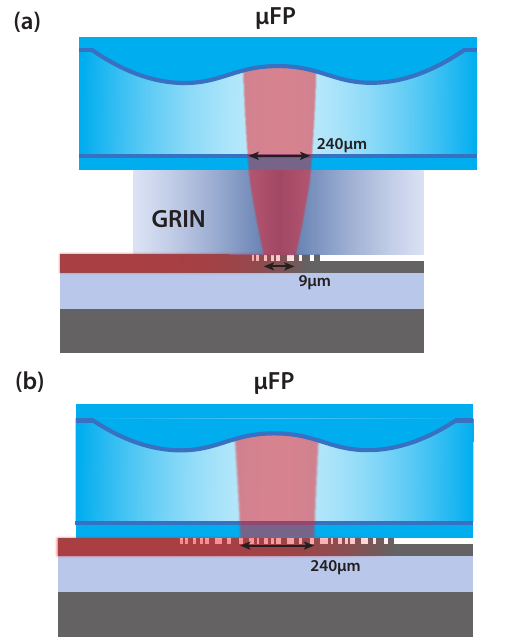}
\caption{Schematic of co-packaged {\textmu}FP with PIC. (a) Small PSGC with GRIN lens to help with mode matching. (b) Numerically optimized large PSGC to match {\textmu}FP mode directly.}
\end{figure}

Replacing the fiber mirror with a fiber Bragg grating cavity (FBGc), we next examine how the reflection response of this Fabry-Pérot resonator is imprinted on the response of this multi-port system. This FBGc is a commercial fiber Bragg defect cavity (Teraxion C56514-0005), that consists of two fiber-Bragg grating mirrors that support a standing-wave mode. The cavity produces a single resonance at a 1550 nm center wavelength having a linewidth of 85~MHz and Q-factor of 2.3 million. Fig. 3(c) displays the reflection, $|S_{11}|^2$, and transmission, $|S_{12}|^2$, measurements produced when this FBGc is coupled to the circuit. 
We see that the reflection response of the resonator is clearly imprinted on the transmission response as required for PDH-locking of a laser to the cavity mode. A transmission efficiency of -5.8 dB is achieved, while the cancellation circuit reduces back-reflection to -15.8~dB, corresponding to a back-reflection a back-reflection suppression ratio of 10 dB.

Next, we couple this circuit to a compact, ultra-high finesse micro-Fabry-Pérot ($\mu$FP) resonator that could serve as a chip-integrated frequency reference for future compact optical clock technologies. This $\mu$FP cavity (see Fig. 3(d)) is one of an array of air-gap resonators that was created using a wafer-scale reflow fabrication process described in Ref. \cite{jin2022micro}. Through independent cavity ring-down spectroscopy measurements, the $\mu$FP of interest was found to have an optical linewidth of 60kHz and an optical Q-factor of 3.2~billion. This resonator was designed to support a relatively large Gaussian-beam diameter of 240~$\mu m$ to reduce the impact of thermal noise imparted by the mirror coating on the cavity mode. As a demonstration of the potential for integration of a $\mu$FP cavity with the chip, the 10.4-micron fiber mode was efficiently coupled to the 240 micron resonator mode using a compact (1mm x 0.5mm) GRIN lens. 

During transmission measurements (see Fig.~3(f)), the laser sweeps through the cavity resonance over a time interval that is much shorter than cavity ringdown time, transforming the resonant response of the cavity into a ringdown response. This type of ringdown response is due to the interference between energy transiently stored in the cavity and the transmitted laser light \cite{savchenkov2007optical,tavernier2010magnesium,trebaol2010ringing,lin2014barium}. 
The transmission response, $|S_{12}|^2$, reveals the ringing cavity response atop a transmission response showing a 7 dB average transmission loss, consistent with a small amount of excess loss produced by fiber-GRIN coupling to the resonator. Back-reflection, $|S_{11}|^2$, of -17 dB seen in Fig.~3(f) corresponds to a back-reflection suppression ratio of 10 dB, demonstrating the suppression of unwanted back-reflections using such reflection cancellation circuits.
As discussed in Section IV, a higher level of integration could be achieved by eliminating the fiber umbilical, and directly bonding the GRIN-cavity assembly to the grating coupler. 

Building on these results, higher grating coupler efficiencies and lower optical losses could be realized by creating a highly reflecting back-plane underneath PSGC.  Moreover, further improvement of the back-reflection suppression ratio could be achieved by using tunable directional couplers and phase shifters, which are readily available in Si photonics foundry processes\cite{shen2017deep}, to tune the parameters of the circuit to obtain a much higher (>30 dB) degree of reflection cancellation.

\section{Discussion and Conclusion}

In conclusion, we have demonstrated a circuit interface to efficiently integrate off-chip components with photonic circuits. 
This circuit interface maps incident and reflected waves to distinct ports of an optical system, permitting direct detection of the reflected wave for PDH locking to the cavity mode\cite{kudelin2023photonic}. The PSGC, a key component of this system, was numerically optimized to realize a peak efficiency of 55\% (2.6~dB), enabling an insertion loss as low as 5.8 dB and a back-reflection suppression ratio of >9 dB using this system. The demonstrated back reflection suppression ratio was limited by the imperfection of the splitting ratio of 50/50 directional coupler. Hence, much higher (>30 dB) back-reflection suppression ratios should be possible with further refinements. For example, the incorporation of a tunable directional coupler and an active phase shifter, would not only permit much higher back reflection suppression but would use such circuit interfaces to efficiently couple to a range of the off-chip photonic components.

Looking beyond Fabry-Pérot resonators, such reflection cancellation circuits can also serve as versatile interface for many other off-chip systems such as cavity optomechanical systems, vapor cells, sensors, and quantum atomic systems. 
Furthermore, we can adapt this system for devices exhibiting polarization-dependent reflection response by substituting the 50/50 directional coupler and balanced passive delay with a tunable coupler and tunable phase shifter. Even when the reflection response is not identical for both polarizations ($\Tilde{R_1}\neq \Tilde{R_2}$), provided $\mu^2=\frac{\Tilde{R_1}}{\Tilde{R_1}+\Tilde{R_2}e^{2i\phi}}$, the reflected wave energy can be directed into Port 2, thereby ensuring a null back reflection for Port 1. This broadens the scope of its application, extending it to more general case of diffuse scattering, as is often obtained in applications such as LIDAR. Note that the cancellation process relies on the presence of two modes in free space, which, in the case above, correspond to two different polarizations. If either $\Tilde{R_1}$ or $\Tilde{R_2}$ were to be zero, it would result in vanishing $|S_{12}|^2$. While the reflection cancellation scheme that we have demonstrated utilizes two polarizations of light, this same concept can be implemented using only a single polarization provided that two spatial modes of the same polarization are used to implement reflection cancellation.

While we utilized a fiber umbilical to couple to off-chip components through this study, a higher degree of integration could be achieved by directly attaching components to the grating coupler. For example, by attaching a GRIN lens to the PSGC (Fig. 4(a)), a compact $\mu$FP of the type used here could be integrated into the system to enable compact new optical clock technologies \cite{kudelin2023photonic}. Moreover, with further computational power, it should be possible to design a PSGC with larger beam spot size that can directly couple to $\mu$FP, without the need of GRIN lens (Fig. 4(b)).

Passive reflection cancellation schemes of the type demonstrated here could eliminate the need for isolators and circulators in many instances, offering a path to simpler and more power-efficient integrated photonic circuits.
Hence, this work could pave the way for heterogeneous integration between integrated photonic circuits and high-finesse Fabry-Pérot resonators that are required for next-generation quantum communications, computation, and time-keeping systems.  

\begin{acknowledgments}
We thank Freek Ruesink and Shai Gertler for helpful technical discussions.
This work was supported by Defense Advanced Research Projects Agency (DARPA) under award number HR0011-22-2-0009 as well as U.S. Department of Energy (DoE) under award number DE-SC0019406 and the National Science Foundation (NSF) under award number 2137740. Any opinions, findings, and conclusions or recommendations expressed in this publication are those of the authors and do not necessarily reflect the views of DARPA, DoE, and NSF.

\end{acknowledgments}

\section*{Data Availability Statement}

The data that support the findings of this study are available from the corresponding author upon reasonable request.

\nocite{*}
\bibliography{MainText}

\begin{thebibliography}{38}
\bibitem [{NIS()}]{NIST_DIS}%
  \BibitemOpen
  \href@noop {} {}\bibinfo {note} {Certain equipment, instruments, software, or materials are identified in this paper in order to specify the experimental procedure adequately. Such identification is not intended to imply recommendation or endorsement of any product or service by NIST, nor is it intended to imply that the materials or equipment identified are necessarily the best available for the purpose.}\BibitemShut {Stop}%
\end{thebibliography}
\end{document}